\RequirePackage[mathlines]{lineno}
\documentclass[amsmath,amssymb,twocolumn,prd,floatfix,showpacs, nofootinbib]{revtex4-2}
\usepackage{tabularx}
\usepackage{bm, color}
\usepackage{overpic,subfigure}
\usepackage{multirow}
\usepackage{array}
\usepackage{dcolumn}
\usepackage[symbol]{footmisc}
\usepackage{epstopdf}
\usepackage{lineno}
\usepackage{hyperref}
\usepackage{amsmath}

\hypersetup{hypertex=true,
	        colorlinks = true,
            linkcolor = blue,
            urlcolor = blue,
            citecolor = blue}

\newcommand{\BESIIIorcid}[1]{\href{https://orcid.org/#1}
   {\hspace*{0.1em}\raisebox{-0.45ex}{\includegraphics[width=1em]{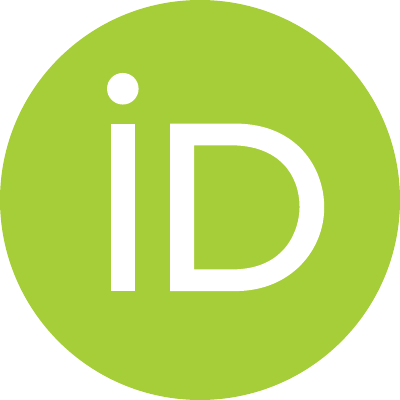}}}
}

\begin{document}
\title{\boldmath Observation of $\psi(3686)\to p K^- K_S^0 \bar \Xi^0+c.c.$ }

\author{M.~Ablikim \textit{et al.}}
\thanks{Full author list given at the end of the Paper.}

\collaboration{BESIII Collaboration}
\noaffiliation

\begin{abstract}
  Using a sample of $(2.712 \pm 0.014) \times 10^{9}$ $\psi(3686)$
  events collected with the BESIII detector, the decay of
  $\psi(3686)\to p K^- K_S^0 \bar \Xi^0+c.c.$ is observed for the
  first time with a statistical significance of $11.5\sigma$. The branching fraction of this decay is measured to be
  $(2.84\pm 0.40\pm 0.25) \times 10^{-6}$, where the first and second
  uncertainties are statistical and systematic, respectively.
This measurement extends the experimental information on 
rare multi-strange $\psi(3686)$ decays and provides an 
experimental reference for future studies of related 
decay modes.
\end{abstract}

\maketitle

\section{Introduction}
The charmonium state $\psi(3686)$ is well described by nonrelativistic dynamics as a bound system of a charm and an anti-charm quark.
In perturbative quantum chromodynamics, its decays are expected to
be dominated by the annihilation of the constituent $c$-$\bar{c}$ into
three gluons or a virtual photon~\cite{ref::tab222}. Radiative
transitions to the $\chi_{cJ}$ states constitute another major decay
pathway with large branching fractions~(BFs).
Most known decays of $\psi(3686)$ contain at most one $s$-$\bar{s}$ quark pair~\cite{PDG}.
Decays involving two strange quark pairs in the final state are
significantly rarer and provide a sensitive probe of the production
mechanisms into the multi-strange hadronic systems. A particularly
interesting class of such processes involves the simultaneous
production of two strange mesons and a double-strange baryon.  The
 $\psi(3686) \to pK^-K_S^0\bar{\Xi}^0$ is such a decay.
Experimental studies of this decay also offer an opportunity to study
strange hadron interactions.
Previously, the decay of $\psi(3686) \to \Lambda(1520)
(\to p K^-) K^- \bar \Xi^+$
has been observed by BESIII, with a BF at the $10^{-6}$ level~\cite{lam1520}.
The channel under study in this paper is related to the $\psi(3686) \to pK^-K^-\bar
{\Xi}^+$ decay by isospin symmetry.  The decay $\psi(3686)\to p K^- K_S^0 \bar
\Xi^0$ should therefore also be detectable at BESIII.

In this paper, we present the first observation and BF measurement of the
$\psi(3686)\to \nobreak p K^- K_S^0 \bar\Xi^0$ decay by analyzing
$(2.712 \pm 0.014) \times 10^{9}$ $\psi(3686)$
events~\cite{psip-num-inc} collected with the BESIII
detector. Throughout this paper, the charge-conjugated mode is always
implied.

\section{BESIII detector and Monte Carlo simulation}
\label{sec:BES}

The BESIII detector~\cite{BESIII:2009fln} records symmetric $e^+e^-$ collisions
provided by the BEPCII storage ring~\cite{Yu:2016cof}
in the center-of-mass energy range from 1.84 to 4.95~GeV,
with a peak luminosity of $1.1 \times 10^{33}\;\text{cm}^{-2}\text{s}^{-1}$
achieved at the center-of-mass of $\sqrt{s} = 3.773\;\text{GeV}$.
The cylindrical core of the BESIII detector covers 93\% of the full solid angle and consists of a helium-based
 multilayer drift chamber~(MDC), a time-of-flight
system~(TOF), and a CsI(Tl) electromagnetic calorimeter~(EMC),
which are all enclosed in a superconducting solenoidal magnet
providing a 1.0~T magnetic field.
The solenoid is supported by an
octagonal flux-return yoke with resistive plate counter muon
identification modules interleaved with steel. The charged-particle momentum resolution at $1~{\rm GeV}/c$ is $0.5\%$,
and the ${\rm d}E/{\rm d}x$ resolution is $6\%$ for electrons
from Bhabha scattering. The EMC measures photon energies with a resolution of $2.5\%$ ($5\%$) at $1$~GeV in the barrel
(end cap) region. The time resolution in the plastic scintillator TOF barrel region is 68~ps, while that in the end cap region was 110~ps.
The end cap TOF system was upgraded in 2015 using multigap resistive plate chamber technology, providing a time resolution of 60~ps~\cite{Li:2017jpg, Guo:2017sjt,  Cao:2020ibk}.
Around 83\% of the data used
in this analysis benefits from this upgrade.

 Monte Carlo (MC) simulated data samples produced with a {\sc geant4}~\cite{GEANT4:2002zbu} based software package,
which includes the geometric description of the BESIII detector and the detector response, are used to optimize the event selection criteria,
 estimate the signal efficiency, and the background level. The simulations incorporate the beam-energy spread and initial-state radiation in the
 $e^+e^-$ annihilation using the generator {\sc kkmc}~\cite{Jadach:2000ir}. The inclusive MC sample includes the production of the
  $\psi(3686)$ resonance, the initial-state radiation production of the $J/\psi$ meson, and the continuum processes incorporated
  in {\sc kkmc}~\cite{Jadach:2000ir}. Particle decays are generated by {\sc evtgen}~\cite{Lange:2001uf, Ping:2008zz} for the known decay
  modes with BFs taken from the Particle Data Group~(PDG)~\cite{PDG} and {\sc lundcharm}~\cite{Chen:2000tv, Yang:2014vra} for the remaining unknown
  ones. Final-state radiation from charged final-state particles is included using the {\sc photos} package~\cite{Richter-Was:1992hxq}.
  The inclusive $\psi(3686)$ MC sample, consisting of 2.712 billion $\psi(3686)$ events, is analyzed with a generic event-type
  examination tool, TopoAna~\cite{Zhou:2020ksj}, to identify the potential backgrounds.

  A mixed signal MC sample of 0.5~M exclusive events is generated. It
  combines the four-body processes
  $\psi(3686)\to p K^-K_S^0 \bar \Xi^0$ and
  \mbox{$\psi(3686) \to \Lambda(1520)(\to p K^-)K_S^0 \bar\Xi^0$} with
  a ratio of approximately $4:1$. which is determined by examining the
  $pK^-$ mass spectrum. The involved processes,
  $\psi(3686) \to \nobreak p\,K^-\nobreak\,K_S^0\nobreak\,\bar\Xi^0$,
  $\psi(3686)\to\nobreak \Lambda(1520)\nobreak K_S^0
  \nobreak\bar\Xi^0$, $K_S^0 \to \nobreak\pi^+\nobreak\pi^-$,
  $\Lambda(1520) \to \nobreak pK^-$,
  $\bar \Xi^0 \to \nobreak \bar \Lambda \pi^0$,
  $\bar \Lambda \to \bar p \pi^+$ and $\pi^0 \to \gamma \gamma$, are
  generated with the corresponding phase space~(PHSP) models,
  which describe the generic $n$-body decays, averaging over the spins
  of initial and final state particles.  The data sample taken at $\sqrt{s}=$ 3.773 GeV
  with an integrated luminosity of 20.3
  fb$^{-1}$~\cite{Ablikim:2013ntc} is used to estimate the continuum
  background contribution.

\section{Event selection and data Analysis}
Charged tracks detected in the MDC are required to be within a polar angle ($\theta$) range of $|\rm{cos\theta}|<0.93$, where $\theta$ is defined with respect to the $z$-axis,
which is the symmetry axis of the MDC. For charged tracks not originating from $K_S^0$ or $\Lambda$ decays, the distance of closest approach to the interaction point (IP) along the $z$-axis, $|V_{z}|$, must be less than 10\,cm 
and the one in the transverse plane, $|V_{xy}|$, less than 1\,cm .
Particle identification~(PID) for charged tracks combines measurements of the energy deposited in the MDC~(d$E$/d$x$)
and the flight time in the TOF to form likelihoods $\mathcal{L}(h)~(h=p,K,\pi)$ for each hadron $h$ hypothesis. Tracks are identified as $p$ candidates when the proton hypothesis has the greatest likelihood
$\mathcal{L}(p)>\mathcal{L}(K)$ and $\mathcal{L}(p)>\mathcal{L}(\pi)$, while $K^-$ candidates are required to satisfy
$\mathcal{L}(K)>\mathcal{L}(\pi)$.

To reconstruct $\bar{\Lambda}$, all possible track pairs with opposite charges are assigned the mass hypotheses $\bar p \pi^+$.
The $\bar p \pi^+$ trajectories are constrained to originate from a common
vertex by a vertex fit, for which the $\chi^2$ is required
to be less than 200.
The $\bar\Lambda$ candidate is then constrained to originate from the IP.
A second vertex fit is then applied to suppress mis-combinations.
The decay length of the $\bar\Lambda$ candidate must be greater than twice the standard deviation of the vertex resolution.
To minimize wrong assignments, the invariant mass of  $\bar p \pi^+$ is required to be within $1.110$~GeV/$c^2$$<M_{\bar p \pi^+} <1.122$~GeV/$c^2$,
which corresponds to $\pm 3\sigma$ around the fitted $\bar \Lambda$ mass peak.
All $\bar\Lambda$ candidates passing the above selection criteria are retained for the subsequent analysis.

Each $K_{S}^0$ candidate is reconstructed from two oppositely charged tracks satisfying $|V_{z}|<$ 20 cm.
The two charged tracks are assigned
as $\pi^+\pi^-$ without imposing further PID criteria. They are constrained to
originate from a common vertex. The
decay length of the $K^0_S$ candidate is required to be greater than
twice the vertex resolution away from the IP.

Photon candidates are identified using showers
in the EMC. The deposited energy of each shower
must be more than 25~MeV in the barrel region
($|\cos \theta|< 0.80$)  and more than 50~MeV in
the end-cap region ($0.86 <|\cos \theta|< 0.
92$). To exclude showers that originate from
charged tracks, the angle subtended by the EMC
shower and the position of the closest charged
track at the EMC must be greater than 10
degrees as measured from the IP. To suppress
electronic noise and showers unrelated to the
event, the difference between the EMC time and
the event start time is required to be within
[0, 700]\,ns.

The $\pi^0$ candidates are reconstructed via the $\pi^0 \to \gamma \gamma$ decays.
The invariant mass of the photon pair is required to be within (0.115, 0.150) GeV/$c^2$.
To improve the momentum resolution, a mass-constrained (1C) kinematic fit to the nominal $\pi^0$ mass is imposed on all the photon pairs and the $\chi_{\pi^0}^2$ is recorded.
All photon combinations for which the 1C fit converges and meets the mass constraint are considered as
potential $\pi^0$ candidates and retained.

The $\bar \Xi^0$ candidates are reconstructed via the $\bar \Xi^0 \to \bar \Lambda \pi^0$ decay, where the
$\bar \Lambda$ and $\pi^0$ candidates are those passing the above selection criteria.

A five-constraint (5C) kinematic fit,
imposing four-momentum conservation on $e^+e^-\to pK^- K_S^0 \bar \Lambda\pi^0$
and with an additional mass constraint on the $\pi^0\to\gamma\gamma$ decay is applied to the candidate events under the hypothesis of $e^+ e^- \to p K^- K_S^0 \bar \Lambda \pi^0$.
In each event, if more than one combination survives, the one with the smallest  $\chi_{\rm total}^{2}=\chi_{\rm 4C}^{2}+\chi_{\rm \pi^0}^{2}$ is retained, and must satisfy $\chi^2_{\rm total}<50$.
The requirement on $\chi^2_{\rm total}$ is optimized by maximizing $\mathit{S}/\sqrt{\mathit{S}+\mathit{B}}$.
Here, $S$ denotes the number of signal events from the signal MC sample, scaled to the branching fraction as determined in this analysis.
The number of background events $B$ is obtained from the inclusive  $\psi(3686)$ MC sample and scaled to data.

\begin{figure}[htbp]
	\begin{center}
		\hspace*{-0.3cm}
		\includegraphics[width=0.50\textwidth]{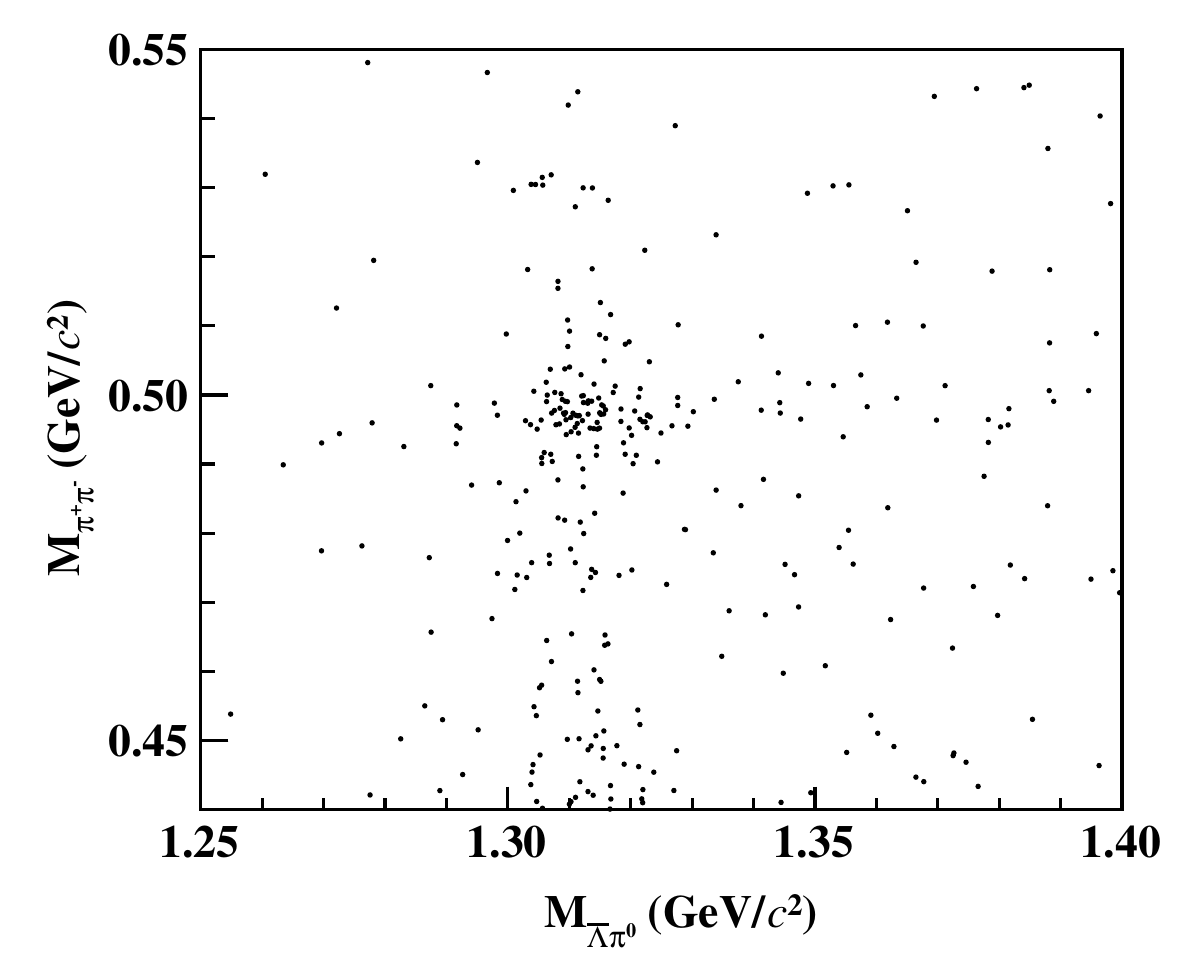}
	\end{center}
  \vspace{-0.8cm}
  \caption{The distribution of $M_{\pi^+\pi^-}$ versus $M_{\bar \Lambda \pi^0}$ for data.
 }
	\label{tab:2D}
\end{figure}

Figure \ref{tab:2D} shows a scatter plot of the invariant masses, $M_{\pi^+\pi^-}$ versus $M_{\bar \Lambda \pi^0}$ for the data.
The cluster around both $K^0_S$ and $\bar \Xi^0$ mass regions indicates the signal of the $\psi(3686)\to p K^-K^0_S\bar \Xi^0$ decay.

\begin{figure*}[htbp]
    \centering
  \scalebox{1.07}
  {
    \includegraphics[width=0.31\textwidth]{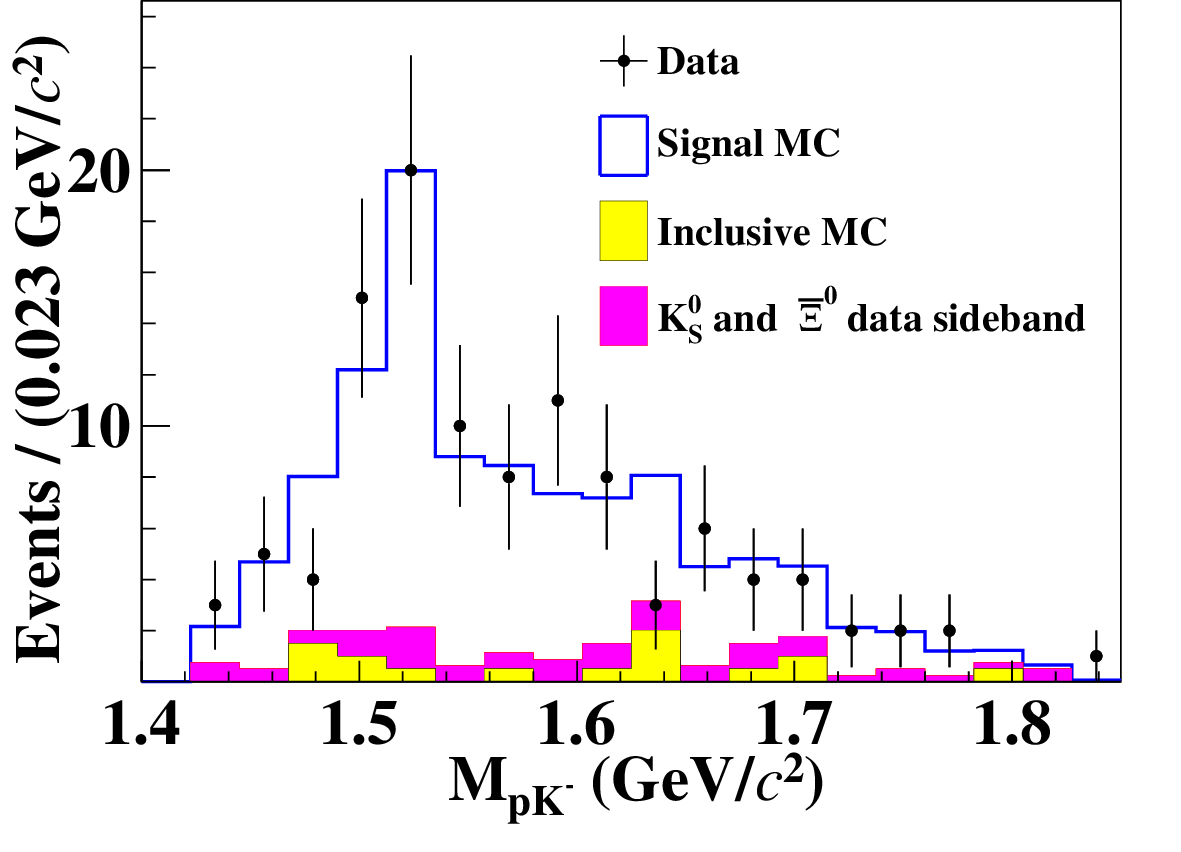}
    \includegraphics[width=0.31\textwidth]{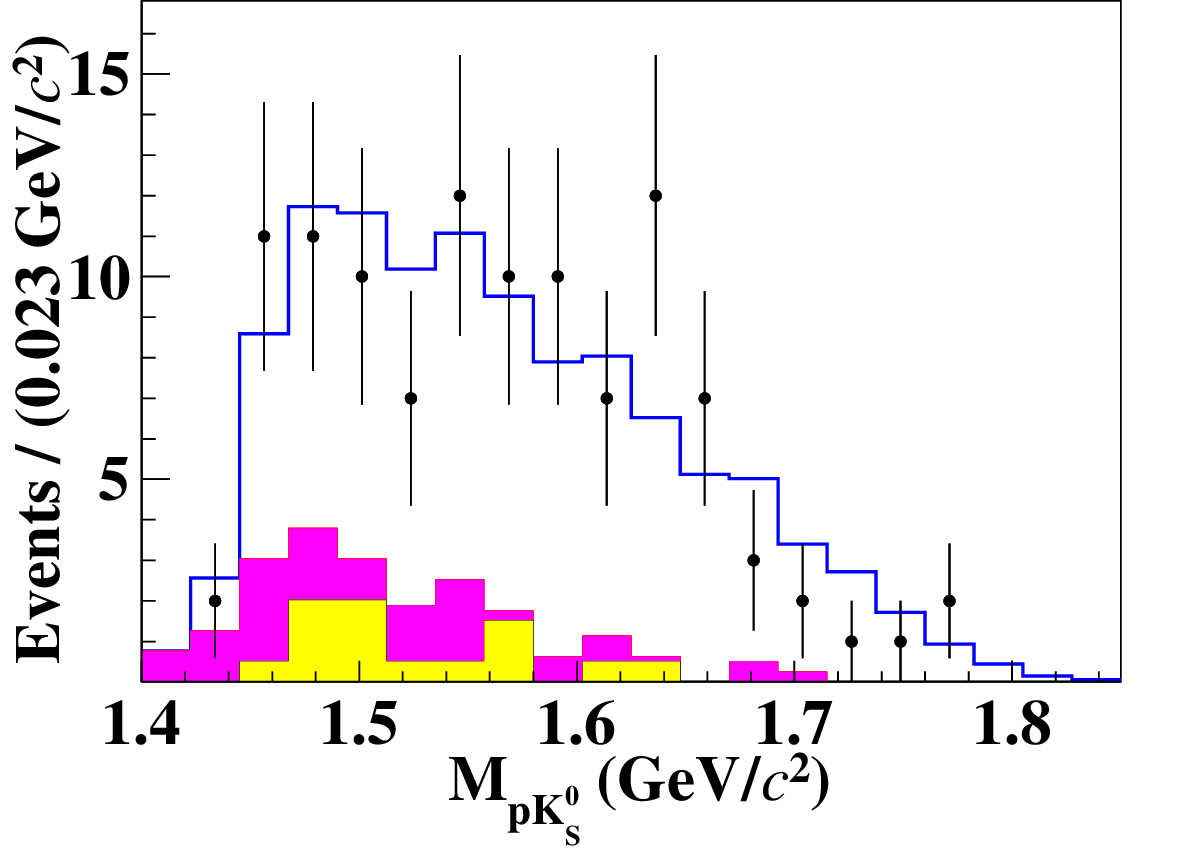}
    \includegraphics[width=0.31\textwidth]{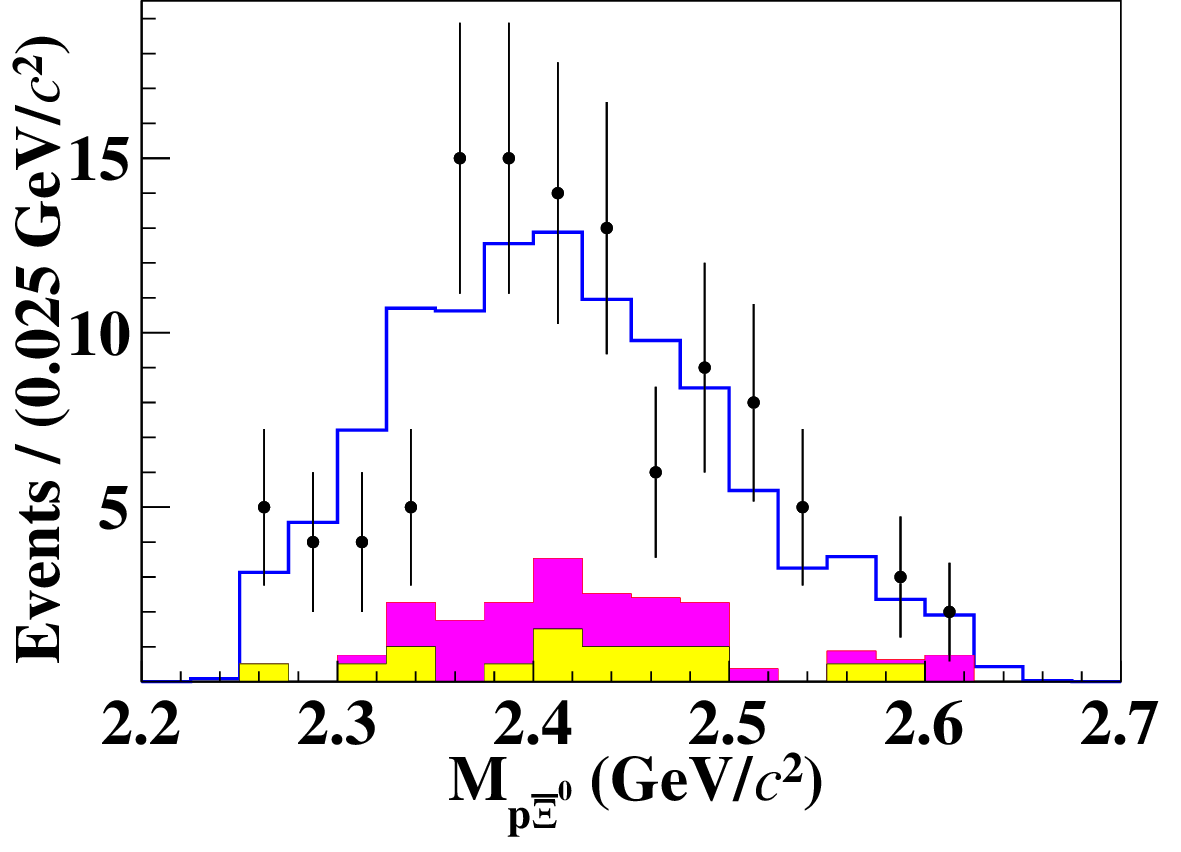}
  }
  \scalebox{1.07}
{
    \includegraphics[width=0.31\textwidth]{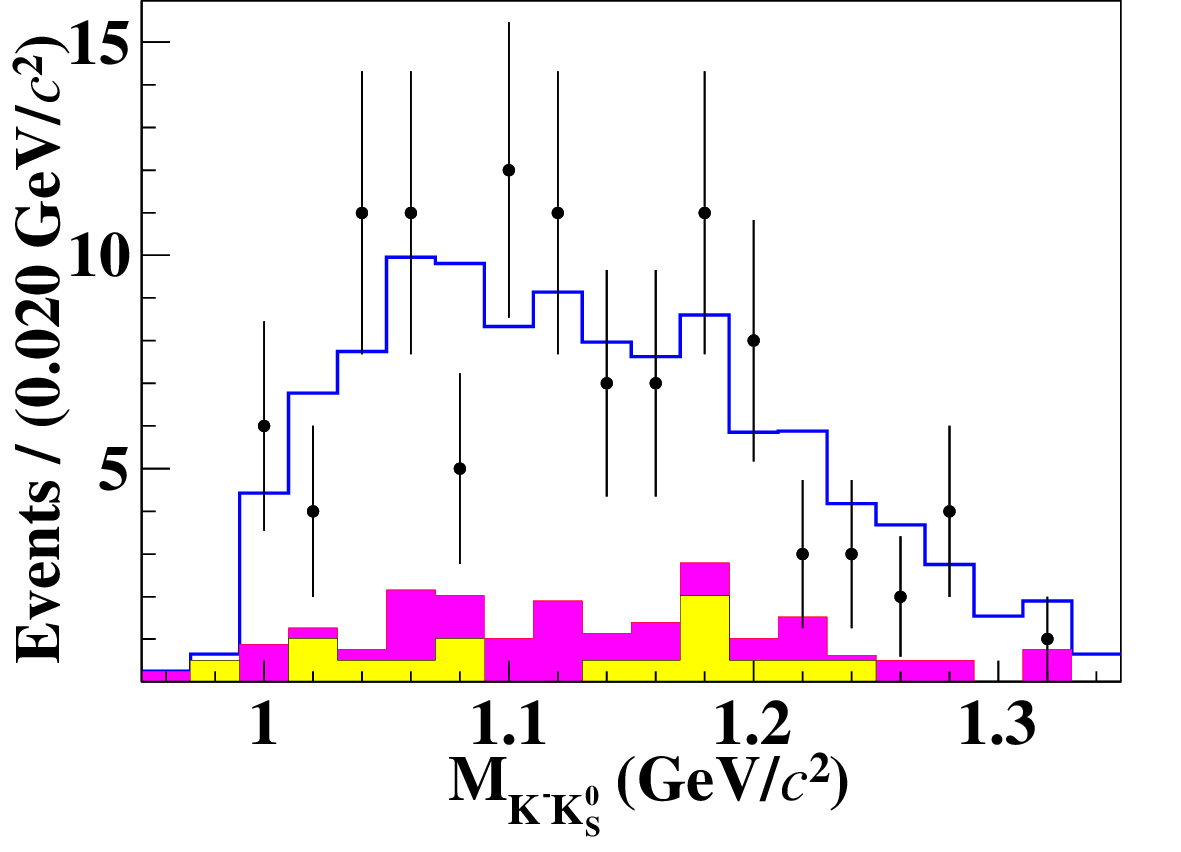}
    \includegraphics[width=0.31\textwidth]{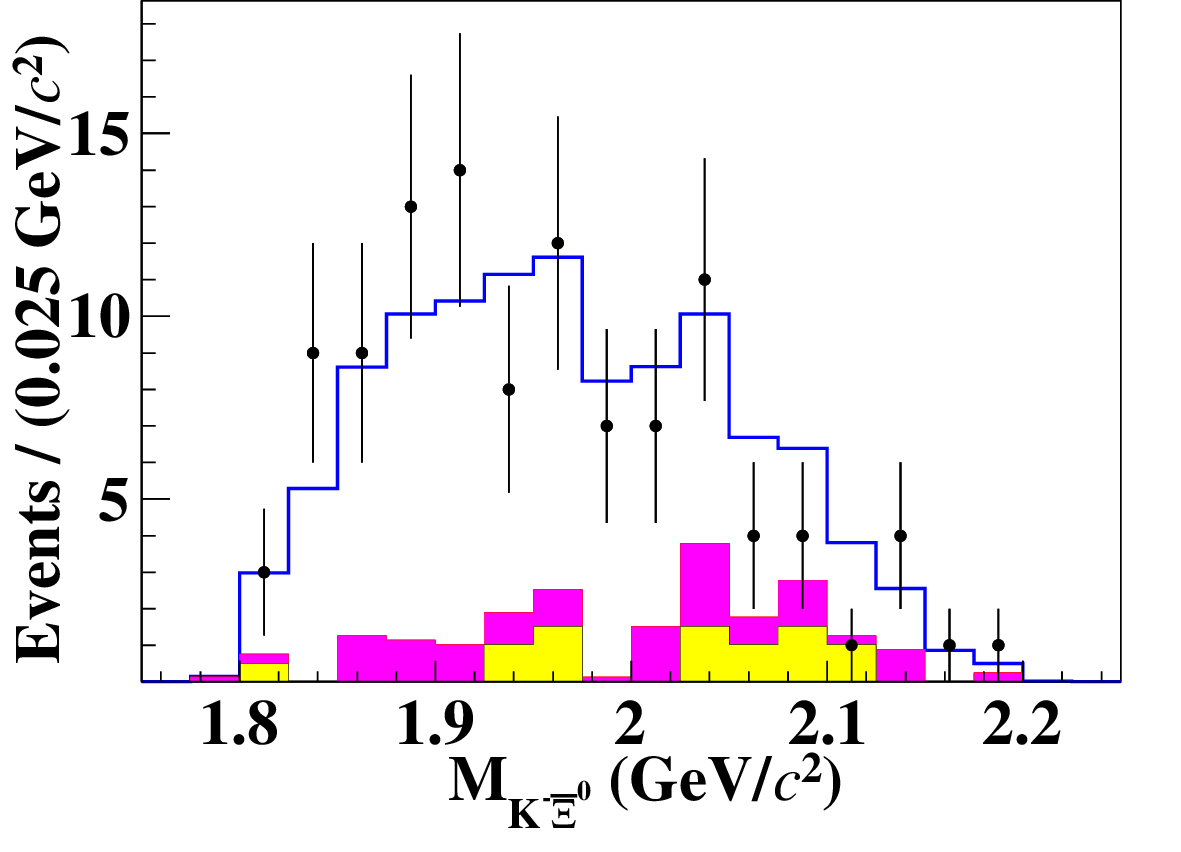}
    \includegraphics[width=0.31\textwidth]{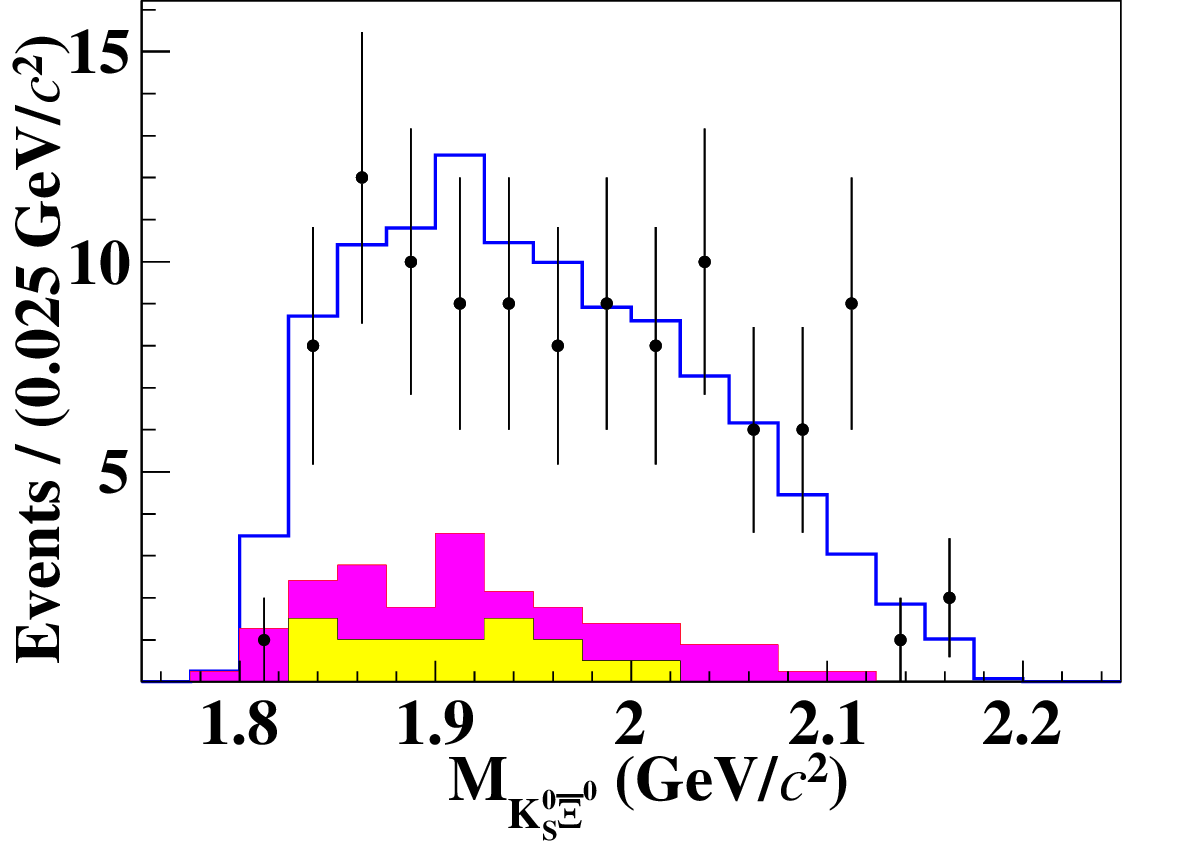}
}
 \vspace{-0.7cm}
 \caption{ Two-body invariant mass distributions of the event
   candidates for $e^+e^- \to pK^- K_S^0 \bar \Xi^0$ in the data and
   the mixed signal MC sample. The dots with error bars denote the
   data at $\sqrt s=3.686$~GeV, and the blue solid lines indicate the
   mixed signal MC sample.  The yellow hatched histograms are the
   normalized backgrounds from the inclusive MC sample and the pink
   hatched histograms are the normalized backgrounds from the $K^0_S$
   and $\bar \Xi^0$ data sideband.  Events have been further required
   to be within the $K_{S}^{0}$ and $\bar \Xi^0$ signal regions
   defined as $M_{\pi^+\pi^-}$ $\in$ $(0.485, 0.509)$~GeV/$c^2$ and
   $M_{\bar \Lambda \pi^0}$ $\in$ $(1.296,1.328)$~GeV/$c^2$,
   respectively; and the $K^0_S$ and $\bar \Xi^0$ sideband regions are
   defined as $M_{\pi^+\pi^-}$ $\in$
   $(0.453,0.474)\cup(0.517,0.541)~{\rm GeV}/c^2$ and
   $M_{\bar \Lambda \pi^0}$ $\in$
   $(1.246,1.278)\cup(1.346,1.378)~{\rm GeV}/c^2$, respectively.  }
    \label{compare:Psi3686}
    \end{figure*}

Figure~\ref{compare:Psi3686} shows the two-body invariant mass distributions of $e^+e^- \to \nobreak p\,K^-\nobreak\,K_S^0\nobreak\,\bar\Xi^0$.
The consistencies of the data and mixed signal MC samples demonstrate the reliability of the mixed signal MC sample.

\begin{figure*}[htbp]
  \begin{center}
  \hspace*{-2.2cm}
  \includegraphics[width=1.23\textwidth]{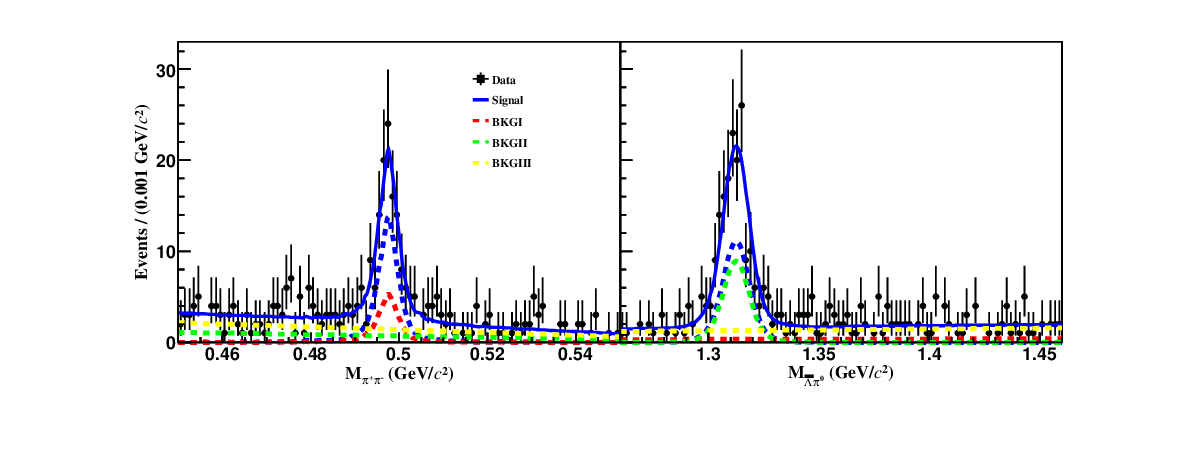}
  \end{center}
  \vspace{-2.2cm}
  \caption{ Projections of the fit onto the $M_{\pi^+\pi^-}$ and
    $M_{\bar{\Lambda}\pi^0}$ distributions for
    $e^+e^- \to pK^- K_S^0 \bar{\Xi}^0$ from the data sample taken at
    $\sqrt{s}=3.686$~GeV.  Data are shown as the dots with error bars
    and the blue solid curves are the fit results.  The red dashed
    line represents the background with $K_S^0$ and non-$\bar \Xi^0$
    while the green dashed line refers to the background with
    $\bar \Xi^0$ and non-$K_S^0$.  The blue dashed line represents the
    signal component and the yellow dashed line refers to the
    combinatorial background with non-$\bar \Xi^0$ and non- $K_S^0$.
  }
  \label{tab:FitData}
\end{figure*}

To determine the signal yield of $\psi(3686) \to pK^- K_S^0 \bar \Xi^0$ in the data, a
2D unbinned maximum likelihood fit is performed on the distribution of
$M_{\pi^+\pi^-}$ versus $M_{\bar \Lambda\pi^0}$ of the accepted candidate events. The corresponding fit projections are shown in Fig.~\ref {tab:FitData}.
Potential background components from $\psi(3686)$ decays are studied by analyzing the inclusive MC sample,
and no significant background containing both $K_S^0$ and  $\Xi^0$ is observed.

In the fit, the probability density functions of the $K_S^0$ and
$\bar{\Xi}^0$ subsystems are uncorrelated, and each one is composed of the
signal and the non-$K_S^0-\bar \Xi^0$ background components. The signal of $K_S^0$~($\bar \Xi^0$) is modeled using the
individual MC-simulated shape convolved with a Gaussian resolution function, whose parameters are determined from the fit to the $\psi(3686)$ data, to
account for the resolution difference between the data and the MC simulation,
$S_{K_S^0}= \rm MC \otimes G(\mu_{1},
\sigma_{1})$~($S_{\bar \Xi^0}=\rm MC \otimes G(\mu_{2}, \sigma_{2})$).
The non-$K_S^0-\bar \Xi^0$ background shapes are described by the linear functions $B_{K_S^0}$ and $B_{\bar \Xi^0}$, where $B_{K_S^0}$~($B_{\bar \Xi^0}$)
is dominated by the decay $\psi(3686)\to p K^- K_S^0 \bar \Lambda \pi^0$
~($\psi(3686)\to p K^- \pi^+ \pi^- \bar \Xi^0$).  There are three such
background components in the fit, constructed as follows:
\begin{itemize}
\item BKGI denotes the contribution from  $S_{K_S^0}  \oplus B_{\bar \Xi^0}$,
\item BKGII denotes the contribution from  $S_{\bar \Xi^0}  \oplus B_{K_S^0}$,
\item BKGIII denotes the contribution from  $B_{K_S^0} \oplus  B_{\bar \Xi^0}$.
\end{itemize}

From this 2D fit, the signal yield of $\psi(3686) \to pK^- K_S^0 \bar \Xi^0$ in the $\psi(3686)$ data
is determined to be $87.1\pm 11.4$, where the uncertainty is statistical only.
The statistical significance of this  signal is found to be 11.5$\sigma$,
as determined by $\Delta(\ln\mathcal{L})=\ln\mathcal{L}_{\max}-\ln\mathcal{L}_{0}$ with
$\Delta ndf=1$.
Here, $\mathcal{L}_{\max}$ and $\mathcal{L}_{0}$ are the maximum likelihoods with and without $S_{K_S^0} \oplus S_{\bar \Xi^0}$
in the fit; $\Delta ndf$ is the change of number of degrees of freedom.
Additive systematic uncertainties are not included in this significance.

\begin{figure*}[htbp]
    \begin{center}
  \hspace*{-2.2cm}
  \includegraphics[width=1.23\textwidth]{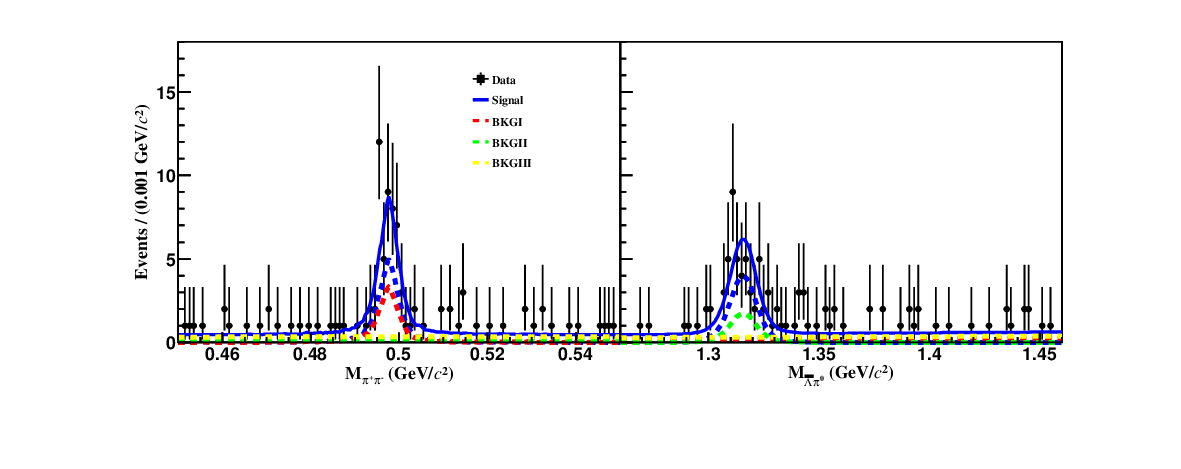}
   \end{center}
  \vspace{-2.2cm}
  \caption{ Projections of the fit onto the $M_{\pi^+\pi^-}$ and
    $M_{\bar {\Lambda}\pi^0}$ distributions for
    $e^+e^- \to pK^- K_S^0 \bar {\Xi}^0$ from the data sample taken at
    $\sqrt{s}=3.773$~GeV.  Data are shown as the dots with error bars
    and the blue solid curves are the fit results.  The red dashed
    line represents the background with $K_S^0$ and non-$\bar \Xi^0$
    while the green dashed line refers to the background with
    $\bar \Xi^0$ and non-$K_S^0$.  The blue dashed line represents the
    signal component and the yellow dashed line refers to the
    combinatorial background with non-$\bar \Xi^0$ and non-$K_S^0$.
  }
        \label{tab:FitQED}
    \end{figure*}

To estimate the continuum background contribution, the data sample taken at $\sqrt s= 3.773~\mathrm{GeV}$ is used.
The same event selection criteria and analysis procedures as used in the analysis of the $\psi(3686)$ data are performed.
In the fit, the signal shape is constructed similarly to that in the $\psi(3686)$ data analysis,
with the resolution parameters of the $K_S^0$ and $\bar{\Xi}^0$  fixed to the values from the fit to the $\psi(3686)$ data.
Figure \ref{tab:FitQED} shows the distributions and the fit results from this data sample.

 \section{Branching fraction}

The BF of $\psi(3686) \to pK^- K_S^0 \bar \Xi^0$ is calculated as
\begin{equation}
  \mathcal{B}(\psi(3686) \to pK^- K_S^0 \bar \Xi^0) = \frac{N_{\text{net}}}{N_{\psi(3686)} \cdot  \prod_{i=1}^{4} \left(\mathcal{B}_i \right) \cdot \epsilon},
\end{equation}
where $N_{\rm net}$ is the net number of the signal yield in the data;
$N_{\psi(3686)}$ is the total number of
$\psi(3686)$ events in data;
$\prod_{i=1}^{4} (\mathcal{B}_i)$ is the product of the BFs of $\mathcal{B}_1 = \mathcal{B}(K_S^0 \to \pi^+\pi^-)=(69.20\pm0.05)\%$, $\mathcal{B}_2 = \mathcal{B}(\bar \Xi^0 \to \bar \Lambda \pi^0)=(99.52\pm0.01)\%$, $\mathcal{B}_3 \nobreak = \nobreak \mathcal{B}(\bar\Lambda \nobreak \to \nobreak \bar p \pi^+)=(64.10\pm0.50)\%$, and $\mathcal{B}_4 = \mathcal{B}(\pi^0 \to \gamma\gamma)=(98.82\pm0.03)\%$, taken from the PDG~\cite{PDG}.
$\epsilon$ is the detection efficiency determined from the signal MC simulation.

The net signal yield of the $\psi(3686) \to pK^-K^0_S\bar \Xi^0$ decay  is calculated with
\begin{equation}
N_{\rm net}= N_{3.686}-N_{\rm continuum,3.686},
\end{equation}
where $N_{\rm continuum,3.686}=N_{\rm continuum,3.773}\times  f_c$;
$N_{3.686}$ and $N_{\rm continuum, 3.773}$ are the fitted
numbers of events in the data samples of 3.686 and 3.773~$\mathrm{GeV}$, respectively. Here, the scale factor of $f_c$ is defined as
\begin{equation}
f_c=\frac{\mathcal L_{3.686}} {\mathcal L_{3.773}}\times\frac{3.773^{2n}}{3.686^{2n}}\times\frac{\epsilon_{\rm continuum,3.686}}{\epsilon_{\rm
continuum,3.773}},
\label{eq:fc}
\end{equation}
where $\mathcal L_{3.686}=3.9~{\rm fb}^{-1}$ and $\mathcal L_{3.773} = 20.3~{\rm fb}^{-1}$ are the integrated luminosities of
the data samples at $\sqrt s=$ 3.686 GeV and 3.773 GeV~\cite{Ablikim:2013ntc};
the detection efficiencies of the continuum production at these two center-of-mass energies are
$\epsilon_{\rm continuum,3.686}=2.17\%$ and $\epsilon_{\rm continuum,3.773}=2.42\%$, respectively.
 The index $n$ is assumed to be 1 for the nominal result, which corresponds
 to a $1/s$ dependence for the input cross section lineshape.
 The scale factor is calculated to be 0.18 and $N_{\rm continuum,3.686}$ is $5.6\pm1.2$.

The measured BF and all relevant numbers used in its calculation are shown in Table~\ref{tab:Branching}.
\begin{table}[htbp]
\centering
\caption{The signal yield in data~($N_{3.686}$), the estimated background yield~($N_{\rm continuum,3.773}$), the detection efficiency~($\epsilon$),
and the determined BF~($\mathcal B$), where the uncertainties are statistical only.}
\begin{tabular}{l|c}
\hline
\hline
  $ N_{3.686}$           & $87.1\pm11.4$  \\ \hline
  $ N_{\rm continnum, 3.773}$           & $31.2\pm6.6$  \\ \hline
  $ N_{\rm continnum, 3.686}$  & $5.6\pm1.2$   \\ \hline
  $ N_{\rm net}$   & $81.5\pm11.5$  \\ \hline
  $\epsilon$                 & $2.43\%$\\ \hline
 $\mathcal{B}$               & $(2.84\pm0.40)\times10^{-6}$\\
  \hline
  \hline
\end{tabular}
    \label{tab:Branching}
\end{table}

The analysis procedure is
validated using 50 statistically independent pseudo-experiments with the same sample size as the real data. The same analysis procedures used for the data are applied to each simulated sample. The average of the measured BFs agrees well with the input BF, indicating no bias in the analysis procedure.

\section{Systematic uncertainty}
Several sources of systematic uncertainties in the BF measurement are evaluated, as discussed below.

The tracking efficiency of protons is studied with the decay of $J/\psi\to p\bar p\pi^+\pi^-$, where a $p$ or $\bar p$ is missed in the sample. The difference in tracking efficiency between data and MC simulation~(denoted as data-MC difference hereafter) is assigned as the systematic uncertainty,
and is $1.0\%$ per track~\cite{ref::tracking}. Using the same control sample and similar analysis method, the systematic uncertainty due to PID of
proton is also assigned to be $1.0\%$ per track~\cite{ref::pid}.

The $K^0_S$ reconstruction efficiency, including the tracking of charged pions, the $K^0_S$ mass window, the vertex fit and the second vertex fit, is estimated using control samples of $J/\psi \rightarrow K^*(892)^{\pm} \bar{K}^\mp$ and $J/\psi \nobreak
\rightarrow \nobreak \phi\nobreak\,K^*(892)^{\pm}\nobreak\,\bar{K}^\mp$.
The data-MC difference in $K^0_S$ reconstruction efficiency is taken as the systematic uncertainty; it is found to be 1.5\% per $K^0_S$~\cite{ref::kso}.

The $\Lambda$ reconstruction efficiency, including the tracking of the
$p\pi^-$ pair, the decay length cut, the $\Lambda$ mass window, the
vertex fit and the second vertex fit, is studied with control
samples of $J/\psi\to pK^-\Lambda$ and
$J/\psi\to \Lambda\bar \Lambda$.  The data-MC difference in $\Lambda$
reconstruction efficiency, 2.0\%, is assigned to be the systematic
uncertainty~\cite{ref::lambdaerror}.

The $\pi^0$ reconstruction efficiency, including the photon selection
and the mass-constrained kinematic fit, is studied by using the
control sample of $D^0\to K^-\pi^+\pi^0$ and $D^0\to K^0_S\pi^0$.  The
data-MC difference in $\pi^0$ reconstruction efficiency is taken as
the systematic uncertainty, 2.0\% per $\pi^0$~\cite{pi0}.

The uncertainty of the 2D fit on the $M_{\pi^+\pi^-}$ versus
$M_{\bar \Lambda\pi^0}$ distribution is considered in three parts. The
first uncertainty due to the signal shapes of $K_S^0$ and $\bar \Xi^0$
is studied by removing the Gaussian resolution functions, and the
change in the fitted signal yield, $1.3\%$, is taken as the systematic
uncertainty.  The second uncertainty due to the background shapes is
estimated by varying the order of the polynomial function, and the
change in the fitted signal yield, $1.5\%$, is taken as the systematic
uncertainty.  The third uncertainty due the fit range in each of
$M_{\pi^+\pi^-}$ and $M_{\bar \Lambda \pi^0}$ is studied by refitting
the data in several of different fit ranges, and the maximum change of
the fitted signal yield, $0.1\%$, is taken as the systematic
uncertainty.  Combining these three parts in quadrature gives the
systematic uncertainty due to the 2D fit to be $2.0\%$.

The uncertainty of the MC modeling is estimated by
comparing the detection efficiency with the nominal signal MC sample and that from an alternative pure PHSP signal MC sample.
The change in the signal efficiency, $2.9\%$, is taken as the systematic uncertainty.

The uncertainty due to the 5C kinematic fit is estimated by comparing the signal efficiencies with and without the helix parameter corrections
with the correction factors from the previous study in Ref.~\cite{ref::helixp}.
The change in the signal efficiency, $1.6\%$, is assigned as the systematic uncertainty.

The uncertainty due to the limited signal MC sample is calculated by $\frac{1}{\sqrt{N}}\sqrt{\frac{1-\epsilon}{\epsilon}}$,
where $\epsilon$ is detection efficiency and $N$ is the produced number of signal MC events.

The intermediate BFs and their uncertainties of $K_S^0 \to \pi^+\pi^-$, $\Lambda \to p\pi^-$, $\bar\Xi^0\to\Lambda\pi^0$,
and $\pi^0\to\gamma\gamma$ are quoted from the PDG~\cite{PDG}.

The uncertainty from the possible interference between the resonance and the continuum amplitudes is evaluated following Ref.~\cite {ref::resonance}.
The cross section for a specific exclusive final state $f$ can be written as
\begin{equation}
\sigma_{\mathrm{total}}^f(s)=\left|a_c^f(s)+e^{i\phi}\cdot a_R^f(s)\right|^2,
\end{equation}
where $a_c^f(s)$ and $a_R^f(s)$ are the continuum and the resonant amplitudes, respectively,
and $\phi$ is their relative phase.
The impact from the interference with respect to the resonance term is defined as
\begin{equation}
\begin{aligned}
  r_R^f = \frac{2}{\hbar c} A B \sin\phi,~ A = \sqrt{\frac{\sigma_c^f(s)}{\mathcal{B}_f}},
\end{aligned}
\end{equation}
where $\hbar c$ is the conversion constant, $\sigma_c^f(s)$ is the cross section of the continuum process, $\mathcal{B}_f$ is
the BF of $\psi(3686) \to pK^-K_S^0 \bar \Xi^0$ obtained in this analysis, and $B$ is a constant depending on the  resonance
parameters quoted from Ref.~\cite {ref::resonance}.
The cross section $\sigma_c^f(s)$ is calculated by
\begin{equation}
  \sigma_c^f(s)=\frac{N_c}{\mathcal{L}_{3.773}\epsilon_{\rm continuum,3.773}}\times\frac{3.773^{2n}}{3.686^{2n}},
\end{equation}
where $N_{c} = 30.1 \pm 6.4$ is the signal yield determined from the
fit to the data at $\sqrt s=3.773$~GeV.  Due to the insufficient amount of
off-resonance data, the relative phase angle can not be
determined.  Considering a uniform prior for the
interference angle, the difference of $r_R^f$ between
$\phi=90^{\circ}$ and $\phi=-90^{\circ}$ is divided by a factor of
$\sqrt {12}$~\cite{BESIII:2021mus}. The resulting value, $6.3\%$, is
conservatively assigned as the systematic uncertainty due to the
possible interference.

The uncertainty due to the scale factor $f_{c}$ is considered in two
parts.  The statistical uncertainty of the MC-determined ratio,
$\frac{\epsilon_{\rm continuum,3.686}}{\epsilon_{\rm
    continuum,3.773}}$= $0.897\pm0.011$, is propagated according to
Eq.~\ref{eq:fc}.  The uncertainty associated with the energy dependence of
the cross section is evaluated by varying $n$ from 1 to 3 according to
the studies in the energy dependence factor $n$ on the cross sections
for $e^+e^-\to light~hadrons$~\cite{ref::f_co}.  Its effect on the BF,
1.0\%, is taken as the corresponding systematic uncertainty.

The total number of $\psi(3686)$ events in data is measured to be $N_{\psi(3686)}=(2.712 \pm 0.014) \times 10^{9}$ with the inclusive hadronic events, as described in Ref.~\cite{psip-num-inc}. The uncertainty of $N_{\rm \psi(3686)}$ is 0.5\%~\cite{psip-num-inc}.

Table \ref{tab:Sys} summarizes the systematic uncertainties in the measurement of the BF of $\psi(3686)\to \nobreak p \nobreak K^-\nobreak  K_S^0 \bar \nobreak \bar \Xi^0$.
The total systematic uncertainty is determined by adding the individual systematic uncertainties in quadrature, assuming they are independent.

\begin{table}[htbp!]
  \centering
   \caption{Sources and assigned relative systematic uncertainties in the BF measurement.}
      \begin{tabular}{l|c}
  \hline
  \hline
 Source                               & Uncertainty~(\%)  \\   \hline
 $p$ and $K^-$ tracking               & 2.0\\  \hline
 $p$ and $K^-$ PID         & 2.0\\  \hline
 $K_{S}^{0}$ reconstruction           & 1.5\\  \hline
 $\Lambda$ reconstruction             & 2.0\\  \hline
 $\pi^0$ reconstruction               & 2.0\\  \hline
 2D fit                               & 2.0\\  \hline
 MC model                             & 2.9\\  \hline
 5C kinematic fit                     & 1.6\\  \hline
 MC statistics                        & 0.9                   \\  \hline
$\mathcal{B}(\bar{K}^0\to K_S^0,K_{S}^{0}\to\pi^+\pi^-)$  & 0.1\\  \hline
$\mathcal{B}(\Lambda \to p\pi^-)$      & 0.8 \\  \hline
$\mathcal{B}(\bar\Xi^0\to\Lambda\pi^0)$  & 0.1\\  \hline
Potential interference         & 6.3\\  \hline
$f_{c}$ factor                           & 1.0\\  \hline
 $N_{\psi(3686)}$                     & 0.5\\  \hline
Total                                     & 8.7  \\
  \hline
  \hline
        \end{tabular}
    \label{tab:Sys}
 \end{table}

\section{Summary}
Using $(2.712 \pm 0.014) \times 10^{9}$ $\psi(3686)$ events taken by 
the BESIII detector, the decay $\psi(3686) \to pK^{-}K_{S}^{0}\bar{\Xi}^{0} + c.c.$
is observed with a statistical significance of $11.5\sigma$.
Its decay BF is determined to be $(2.84 \pm 0.40 \pm 0.25) \times 10^{-6}$, 
where the first and second uncertainties are statistical and 
systematic, respectively.
This is the first measurement of the decay 
$\psi(3686) \to pK^{-}K_{S}^{0}\bar{\Xi}^{0} + c.c.$, containing two strange mesons and one double-strange baryon.
A possible $\Lambda(1520)$ contribution is indicated in the 
$pK^{-}$ invariant mass distribution, and the corresponding 
$\Lambda(1520)K_{S}^{0}\bar{\Xi}^{0}$ channel is related by 
isospin symmetry to the 
$K^{-}\Lambda(1520)\bar{\Xi}^{+}$ channel measured in 
Ref.~\cite{lam1520}.
Further studies with high statistics data taken at the future super
tau-charm factory~\cite{stcf} will be valuable to further understand
the decay mechanisms of these types of decays, and to seek potential
new structures.

\section{Acknowledgement}
The BESIII Collaboration thanks the staff of BEPCII 
(\href{https://lssf.cas.cn/sszs/zyyj/bjzfdzdzj/}
{https://cstr.cn/31109.02.BEPC}) and the IHEP 
computing center for their strong support. This work 
is supported in part by National Key R\&D Program 
of China under Contracts Nos. 2025YFA1613900,
2023YFA1606000, 2023YFA1606704; National Natural 
Science Foundation of China (NSFC) under Contracts 
Nos. 11635010, 11935015, 11935016, 11935018, 
12025502, 12035009, 12035013, 
12061131003, 12192260, 12192261, 12192262, 
12192263, 12192264, 12192265, 12221005, 12225509, 
12235017, 12342502, 12361141819, 12535005; the 
Chinese Academy of Sciences (CAS) Large-Scale 
Scientific Facility Program; the Strategic Priority 
Research Program of Chinese Academy of 
Sciences under Contract No. XDA0480600; CAS under 
Contract No. YSBR-101; 100 Talents Program of CAS; 
The Institute of Nuclear and Particle Physics 
(INPAC) and Shanghai Key Laboratory for Particle 
Physics and Cosmology; Agencia Nacional de 
Investigación y Desarrollo de Chile (ANID), Chile 
under Contract No. ANID CCTVal CIA250027; ERC under 
Contract No. 758462; German Research Foundation DFG 
under Contract No. FOR5327; Istituto 
Nazionale di Fisica Nucleare, Italy; Knut 
and Alice Wallenberg Foundation under Contracts Nos. 
2021.0174, 2021.0299, 2023.0315; Ministry of 
Development of Turkey under Contract No. 
DPT2006K-120470; National Research Foundation of 
Korea under 
Contract No. RS-2026-25486791; 
National Science and Technology fund of Mongolia; 
Polish National Science Centre under Contract No. 
2024/53/B/ST2/00975; STFC (United Kingdom); Swedish 
Research Council under Contract No. 2019.04595; 
U. S. Department of Energy under 
Contract No. DE-FG02-05ER41374.

\nolinenumbers
\onecolumngrid

\centering
\vspace{0.5cm}
\noindent\rule{0.5\textwidth}{1pt}
\vspace{0.5cm}

M.~Ablikim$^{1}$\BESIIIorcid{0000-0002-3935-619X},
M.~N.~Achasov$^{4,c}$\BESIIIorcid{0000-0002-9400-8622},
P.~Adlarson$^{81}$\BESIIIorcid{0000-0001-6280-3851},
X.~C.~Ai$^{87}$\BESIIIorcid{0000-0003-3856-2415},
C.~S.~Akondi$^{31A,31B}$\BESIIIorcid{0000-0001-6303-5217},
R.~Aliberti$^{39}$\BESIIIorcid{0000-0003-3500-4012},
A.~Amoroso$^{80A,80C}$\BESIIIorcid{0000-0002-3095-8610},
Q.~An$^{77,64,\dagger}$,
Y.~H.~An$^{87}$\BESIIIorcid{0009-0008-3419-0849},
Y.~Bai$^{62}$\BESIIIorcid{0000-0001-6593-5665}
% ...
\\
\vspace{0.2cm}
(BESIII Collaboration)\\
\vspace{0.2cm} {\it
$^{1}$ Institute of High Energy Physics, Beijing 100049, People's Republic of China\\
% ...
$^{4}$ Budker Institute of Nuclear Physics SB RAS (BINP), Novosibirsk 630090, Russia\\
$^{5}$ Carnegie Mellon University, Pittsburgh, Pennsylvania 15213, USA\\
% ...
$^{31}$ INFN Sezione di Ferrara, (A)INFN Sezione di Ferrara, I-44122, Ferrara, Italy; (B)University of Ferrara, I-44122, Ferrara, Italy\\
% ...
$^{39}$ Johannes Gutenberg University of Mainz, Johann-Joachim-Becher-Weg 45, D-55099 Mainz, Germany\\
% ...
$^{62}$ Southeast University, Nanjing 211100, People's Republic of China\\
% ...
$^{64}$ State Key Laboratory of Particle Detection and Electronics, Beijing 100049, Hefei 230026, People's Republic of China\\
% ...
$^{77}$ University of Science and Technology of China, Hefei 230026, People's Republic of China\\
% ...
$^{80}$ University of Turin and INFN, (A)University of Turin, I-10125, Turin, Italy; (B)University of Eastern Piedmont, I-15121, Alessandria, Italy; (C)INFN, I-10125, Turin, Italy\\
$^{81}$ Uppsala University, Box 516, SE-75120 Uppsala, Sweden\\
% ...
$^{87}$ Zhengzhou University, Zhengzhou 450001, People's Republic of China\\
\vspace{0.2cm}
$^{\dagger}$ Deceased\\
% ...
$^{c}$ Also at the Novosibirsk State University, Novosibirsk, 630090, Russia\\
% ...
}

\end{document}